\documentclass{ifacconf}

\usepackage{graphicx}      
\usepackage{natbib}        
\usepackage{amsmath,amssymb}
\usepackage{colortbl}
\usepackage{xcolor}
\usepackage{tikz}
\usetikzlibrary{positioning, shapes, arrows, calc}
\usetikzlibrary{arrows.meta, positioning}
\usepackage{adjustbox}
\usepackage{pgfplots}
\usepackage[bb=boondox]{mathalfa}
\usepackage{comment}
\usepackage{booktabs}
\usepackage{subfigure}

\newtheorem{assumption}{Assumption}
\newtheorem{design}{Design constraint}
\newtheorem{lemma}{Lemma}

\newtheorem{proposition}{Proposition}

\newtheorem{example}{Example}
\newtheorem{remark}{Remark}
\newtheorem{policy}{Policy}

\begin{document}
\begin{frontmatter}
\title{Optimal policy design for innovation diffusion: shaping today's incentives for transforming the future}

\thanks[footnoteinfo]{This work was supported by the European Union – Next Generation EU, Mission 4, Component 1, through the PRIN Project TECHIE: “A Control and Network-Based Approach for Fostering the Adoption of New Technologies in the Ecological Transition” (CUP Master: D53D23001320006, CUP: B53D23002760006) under Grant 2022KPHA24.}

\author[First]{Lisa Piccinin} 
\author[Second]{Valentina Breschi} 
\author[Third]{Chiara Ravazzi}
\author[Third]{Fabrizio Dabbene}
\author[First]{Mara Tanelli}

\address[First]{Politecnico di Milano, 20133,
Milan, 
Italy (e-mail: lisa.piccinin@polimi.it, mara.tanelli@polimi.it).}
\address[Second]{Eindhoven University of Technology,
5600 MB, Eindhoven, The Netherlands (e-mail: v.breschi@tue.nl)}
\address[Third]{Institute of
Electronics, Computer and Telecommunication Engineering, National
Research Council of Italy (CNR-IEIIT), 10129, Turin, Italy (e-mail: chiara.ravazzi@cnr.it, fabrizio.dabbene@cnr.it)}

\begin{abstract}                 
In this paper, we propose a new framework for the design of incentives aimed at promoting innovation diffusion in social influence networks. In particular, our framework relies on an extension of the Friedkin and Johnsen opinion dynamics model characterizing the effects of $(i)$ short-memory incentives, which have an immediate yet transient impact, and $(ii)$ long-term structural incentives, whose impact persists via an exponentially decaying memory. We propose to design these incentives via a model-predictive control (MPC) scheme over an augmented state that captures the memory in our opinion dynamics model, yielding a convex quadratic program with linear constraints. Our numerical simulations based on data on sustainable mobility habits show the effectiveness of the proposed approach, which balances large-scale adoption and resource allocation. 
\end{abstract}


\end{frontmatter}

\section{Introduction} 
Given the urge to address climate change, promoting the diffusion of sustainable behaviors is a key societal challenge that hinges on a shift in collective attitude. Such a change in collective behavior is often hampered by individual resistance to such sustainable behaviors (e.g., the adoption of green technologies), which are shaped (among others) by homophily, but can be significantly accelerated by strategic external incentives. Specifically, policymakers and stakeholders can rely on two main types of incentives to encourage adoption. On the one hand, \textit{on--off} incentives (e.g., first-time user discounts and temporary promotions) can promote the use of a service and temporarily boost its usage. Nonetheless, this type of incentive only has a short-term impact on individuals' habits~\citep{c3}. 
On the other hand, \textit{systemic} policies (e.g., building dedicated infrastructure or educational campaigns) can progressively foster long-term acceptance leading to behavioral shifts~\citep{c20}. At the same time, they often require considerable economic efforts with no guarantees of effectiveness. 

In this setting, combining opinion dynamics and control theory offers a systematic way to analyze how opinions evolve and to design incentives. Understanding the dynamics governing opinion formation in social networks has been a persistent goal in opinion dynamics theory, and several models have been proposed over the years (see the recent review in~\cite{anderson2019recent}). One of the most recognized is the DeGroot model~\citep{degroot1974reaching}, which describes how individual opinions evolve under social pressure as a weighted averaging process. Specifically, people update their beliefs by combining their own opinion with those with whom they interact. This seminal model has then been extended to include other aspects of opinion formation, with the most famous example being the Friedkin-Johnsen (FJ) model~\citep{friedkin1990social}. This model introduces the concept of stubbornness, thus also modeling individual resistance to adoption irrespective of external influences. Other frameworks include the Altafini model~\citep{altafini2012consensus} and the Hegselmann–Krause model~\citep{hegselmann2015opinion}, respectively incorporating antagonistic interactions through negative edge weights in the social network and bounded confidence intervals to capture homophily.  

While these (classical) models effectively characterize open-loop opinion dynamics, they inherently assume that no external interventions influence opinion formation apart from social interactions. However, this is not truly the case in practice, as one believes are also shaped by stimuli that are external to social influences (e.g., incentive schemes nudging behavioral shifts). This limitation demanded new models~\citep{sirbu2016opinion}, allowing one to also characterize behavioural changes in response to external stimuli. \cite{ancona2022model} extends the FJ framework with an additional virtual node representing a policymaker, thus explicitly accounting for external interventions. Based on this model,~\cite{ancona2022model} studies how campaigns can strategically exploit network topology and individual susceptibilities to nudge opinions. Meanwhile,~\cite{wendt2019control} presents a DeGroot-based model with a \textit{selfish} agent that updates its opinion by applying proportional and proportional–integral controllers to steer network opinions toward a desired value.

While characterizing the impact of external stimuli, all the previous works do not use the opinion dynamics model they proposed for policy design. Nonetheless, the expressive power of opinion dynamics models can also be leveraged to design (optimal) interventions in a systematic way. This limitation is overcome by some recent works in the literature. For example, in~\cite{gupta2021influencing}, individuals update their opinions under the effect of influence policies, showing that the optimal strategy depends on how likely individuals are to change their opinions.
Instead,~\cite{fontan2024influencing} extends the FJ model by introducing a \textit{policymaker} node that affects agents’ opinions. This work further analyzes how the policymaker can influence collective opinions toward a desired target through repeated campaigns, optimally allocating the budget to minimize the network’s deviation from the target. Meanwhile, in~\cite{moruarescu2018space}, opinions evolve according to the continuous-time DeGroot averaging process and the effect of marketing campaigns is modeled as impulsive updates. In this framework, interventions are designed to control the trade-off between social imitation and personal inclination in order to reach a desired goal.

Despite looking into policy design, all the previous works propose strategies to plan optimal policies offline. However, this choice would not enable policy interventions to adapt to changes in the budget (e.g., if the budget is decremental) and the current policy's effectiveness. This would instead be possible by using Model Predictive Control (MPC) approach, which allows the design of policies at each step based on the current network state and updated budget availability. This strategy is employed in~\cite{ravazzi2025optimal}, where the short-term effect of external interventions is considered for an extension of the FJ model and MPC is employed to design optimal incentives. Nonetheless, this work focuses on the design of \textit{on--off} policies, thus only modelling the effect of short-term incentives. Meanwhile,~\cite{piccinin2025innovation} only accounts for systemic policies in their extension of the FJ model, describing the impact of such policies as integral inputs, thus (simplistically) assuming that their effect accumulates linearly over time. 

\emph{Contribution:} As reported in~\cite{callander2017dynamic}, changes in technology and social contexts reduce the long-term effectiveness of policies, as past interventions can become obsolete. Differently from what is assumed in~\cite{piccinin2025innovation}, this implies that the impact of systemic policies decays over time, even if more slowly than short-term \textit{on--off} interventions. To cope with this limitation, in this work, we propose an opinion dynamics model that includes the decaying effects of long-term interventions, while accounting also for the short-term ones. Similarly to~\cite{ravazzi2025optimal} and~\cite{piccinin2025innovation}, we use the proposed model to design optimal fostering policies balancing the widespread acceptance of the target technology/service and the investment made to reach it under budget constraints. We show this trade-off via some numerical results, providing a sensitivity analysis with respect to the available budget and a set of key model parameters. 

\emph{Outline:} Section~\ref{sec:model} presents the proposed opinion dynamics model, and its properties are analyzed in Section~\ref{sec:properties}. In Section~\ref{sec:control} the formulated control problem to shift agents’ opinions towards the desired state is presented. A numerical example that analyzes the impact of the policies is provided in Section~\ref{sec:num_example}. Finally, Section~\ref{sec:conclusions} summarizes the proposed framework and discusses future developments.

\emph{Notation:} We indicate with $\mathbb{N}$, $\mathbb{N}_{0}$ and $\mathbb{R}_{+}$, the set of natural numbers, the set of natural numbers including zero, and the set of positive real numbers, respectively. Given any vector $ x \in \mathbb{R}^n $ and matrix $ A \in \mathbb{R}^{m \times n}$, their transposes are denoted as $x^{\top}$ and $A^{\top}$, respectively, and the inverse of $B \in \mathbb{R}^{n \times n}$ is given by $B^{-1}$. 
The positive (non-negative) definite matrix $A$ is denoted as $A \succ 0$ ($A \succeq 0$). Moreover, $\|x\|_{2}$ and $\|A\|_{2}$ indicate their 2-norms, $\|x\|_{B}^{2}=x^{\top} Bx$, $x_i \in \mathbb{R}$ denotes the $i$-th component of $x$ and $A_{ij} \in \mathbb{R}$ represents the element of $A$ in position $(i,j)$. The Hadamard product is denoted by the symbol $\odot$. We denote identity matrices with $I$, we indicate the vectors of ones and zeros (of appropriate dimensions) with $\mathbb{1}$ and $\mathbb{0}$, respectively. 

\section{FJ Model with short and long-term incentives}\label{sec:model}
Consider a set $\mathcal{V}$ of  $N \in \mathbb{N}$ agents, all interacting within a social system. Assume that each agent $v \in \mathcal{V}$ is endowed with a scalar value \( x_v(t) \in [0,1] \) at each time \( t \in\mathbb{N}_0\), which measures its probability to be inclined to adopt a new technology/service or to engage in a specific behavior of interest. Let $x(t) \in [0,1]^{N}$ be the column vector stacking these indicators of individual inclination. As in the classical Friedkin and Johnsen model \citep{friedkin1990social}, we assume that $x(t)$ evolves over time as a consequence of $(i)$ mutual interactions within a social system, and $(ii)$ individual initial predispositions or 
other conditions that are not dependent on the social interactions and act as persistent biases over time.

To characterize mutual interactions, let us represent the influence network guiding opinion dynamics as a directed, weighted graph $\mathcal{G} = (\mathcal{V}, \mathcal{E}, P)$, where   $\mathcal{E}$ denotes the set of directed edges encoding social connections between agents\footnote{Agent $v$ influences agent $w$ if $(w,v) \in \mathcal{E}$.}, and $P$ is a nonnegative row-stochastic matrix that captures the strength of mutual influence, satisfying
\begin{equation}\label{eq:properties_P}
    P_{wv}>0,~\forall (w,v) \in \mathcal{E}, \qquad \sum_{v \in \mathcal{V}}  P_{wv}  =  1,~\forall w \in \mathcal{V}.
\end{equation} 
In addition, let $\lambda_i \in [0,1]$ capture the susceptibility of the $i$-th agent's opinion to the influence of its peers, with $\lambda_i=0$ implying that the $i$-th agent’s opinion evolution is driven mainly by its initial belief and eventual external incentives, while $\lambda_i$ close to $1$ indicates that $i$-th agent is strongly influenced by its peers, for all $i \in [1,N]$. We assume that the agents' susceptibilities, encoded in the diagonal matrix $\Lambda \in [0,1]^{N \times N}$, satisfy the following. 
\begin{assumption} \label{ass:connected}
For every node $w \in V$, there exists a path from it to a node $v \in V$ such that $\lambda_{v}<1$. 
\end{assumption}
Apart from all the previous elements that characterize mutual interactions, we further assume that each agent is characterized by a time-varying \textit{inherent bias} $u_v(t) \in [0,1]$, in turn shaped by fostering policies and a persistent \textit{bias} $u^o_v \in [0,1]$ that represents the initial stubbornness\footnote{The closer $u^o_v$ is to $1$, the more predisposed the agent is to adopt or accept a new technology.} of the $v$-th agent, for $v \in \mathcal{V}$. Accordingly, the dynamics of this state is governed by the following equation:
\begin{equation}\label{eq:dyn_model}
    x(t+1) = \Lambda P x(t) + (I-\Lambda) u(t),
\end{equation}
where $u(t) \in [0,1]^{N}$ is the vector stacking the inherent bias of all $v \in \mathcal{V}$. Note that, in the absence of policy intervention or any other factor other than the persistent bias\footnote{$u^{o} \in [0,1]^{N}$ stacks $u_v^o$ for all $v \in \mathcal{V}$.} $u^{o} \in [0,1]^{N}$, \eqref{eq:dyn_model} simplifies to 
\begin{equation}
x(t+1) = \Lambda P x(t) + (I-\Lambda) u^o.
\end{equation} 
In this case, the inclinations of individuals converge to a final limit profile \textcolor{black}{that coincides with the one of the original FJ model}. 
\begin{lemma}[\citep{c17}]\label{lemma:steady_free} Let Assumption~\ref{ass:connected} hold, $x(0) \in [0,1]^{N}$. Then, the $\{x(t)\}_{t\in\mathbb{N}_0}$ converges to
\begin{equation}\label{eq:steady_state_free}
x_{\infty} \!:= \!\lim_{t \to \infty} x(t) 
= (I - \Lambda P)^{-1} (I - \Lambda) u^{\mathrm{o}}.
\end{equation}
\end{lemma}

\subsection{Dual-incentive framework to nudge virtuous behaviors}
In this work, we aim at characterizing the effects of short and long-term incentives on the \emph{inherent bias} $u(t)$ in \eqref{eq:dyn_model}, yet (differently from~\cite{piccinin2025innovation}) considering that past incentives might not have the same influence as the current ones in shaping individual opinions and behaviors.
To this end, we formalize the evolution of $u(t)$ as the result of a combination of both long-term and short-term incentives as follows:
\begin{equation}\label{eq:input_eq}
    u(t) = u^o + \rho \odot u^{\ell, \text{mem}}(t) + (\mathbb{1} - \rho) \odot u^s(t),
\end{equation}
where $\rho \in [0,1]^N$ weights the relative influence of the two kinds of incentives, $u^{\ell, \text{mem}}(t) \in [0,1]^N$ encodes the normalized effect of long-term policies, and $u^s(t) \in [0,1]^N$ represents the short-term component. 
In particular: 
\begin{enumerate}\item[(i)] the short-term component $u^s(t)$ models temporary, \textit{on--off} interventions such as bonuses or promotional campaigns that influence behavior only at the moment of their activation. These incentives trigger immediate service usage without persistently altering agents’ underlying preferences and their effect vanishes once the intervention is removed;
\item[(ii)] the long-term component $u^{\ell, \text{mem}}(t)$ accounts for incentives that persist over time and continue to shape behavior even after their initial application, such as regulatory frameworks, infrastructure investments that influence the adoption's dynamics over an extended period.
\end{enumerate}
The long-term component $u^{\ell, \text{mem}}(t)$ is here modelled through a decaying memory, namely
\begin{equation}\label{eq:mem_input}
    u^{\ell, \text{mem}}(t)\!=\!\sum_{j=0}^{{\color{black}J}} \omega_j u^\ell(t - j - 1), 
\end{equation}
where $J \in \mathbb{N}$ characterizes the memory window from which the impact of a previous nudging intervention vanishes and the weight $\omega_j$ determines the rate at which past incentives lose relevance, satisfying the following assumption.
{\color{black}{The parameter $J$ can be specified according to the modeling framework or empirical context\footnote{Throughout the remainder of the paper, we set 
$J = t -1 $, thereby considering the influence of past incentives on the entire past horizon.}.}}
\begin{assumption}
 \label{ass:weights}
The sequence of weights $\{\omega_j\}_{j \in \mathbb{N}_0}$ are such that 
\begin{equation}\label{eq:weight_prop}
\omega_j\ge 0,\quad \sum_{j=0}^{J}\omega_j\textcolor{black}{\le}1,\quad (\omega_{j+1}\le \omega_j).
\end{equation}
\end{assumption}
While several choices can be made that satisfy the previous assumption, here we introduce two possible options for $\{\omega_j\}_{j \in \mathbb{N}_0}$ to describe a decaying memory.  
\begin{example}[Exponential IIR kernel] \label{ex:IIR}
Let $\kappa\in(0,1)$ and define
\begin{equation}\label{eq:IIR_singleweight}
\omega_j=(1-\kappa)\,\kappa^{\,j},~~~ \kappa=e^{-1/\tau}.
\end{equation}
Then
$$
u^{\ell,\mathrm{mem}}(t)=\sum_{j=0}^{t-1}\omega_j\,u^\ell(t-j -1)
$$
admits the following di expression:
$$
u^{\ell,\mathrm{mem}}(t+1)=\kappa\,u^{\ell,\mathrm{mem}}(t)+(1-\kappa)\,u^\ell(t).$$
Moreover, if $u_{\ell}(t)=0$ for $t\geq T_{\mathrm{off}}$, i.e., the control is switched off at $T_{\mathrm{off}}$, then for $m\geq 0$ steps after switch-off,
$$
u^{\ell,\mathrm{mem}}(T_{\mathrm{off}}+m)=\kappa^{\,m}\,u^{\ell,\mathrm{mem}}(T_{\mathrm{off}}).
$$
\end{example}
\begin{example}[Exponential FIR kernel]
Let $J\in\mathbb{N}$ and $\kappa\in(0,1)$. Define
$$
\omega_j=\frac{(1-\kappa)\,\kappa^{\,j}}{1-\kappa^{\textcolor{black}{J}+1}},
$$
which are nonnegative, monotonically decreasing weights, satisfying $$\sum_{j=0}^{\textcolor{black}{J}}\omega_j=1.$$
For $m=0,\dots,J$, we thus have
$$
u^{\ell,\mathrm{mem}}(T_{\mathrm{off}}+m)
=\frac{(1-\kappa)\kappa^{m}}{1-\kappa^{\textcolor{black}{J}+1}}
\sum_{r=0}^{\textcolor{black}{J}-m}\kappa^{r}\,u^\ell(T_{\mathrm{off}}-r).
$$
Moreover, by nonnegativity of the weights, we get the following lower bound
$$
u^{\ell,\mathrm{mem}}(T_{\mathrm{off}}+m)\ \geq \omega_0\,\kappa^{\,m}u^\ell(T_{\mathrm{off}}),
\quad 0\leq m\leq J,$$
where the term on the right-hand side corresponds to the case in which $m=J$. Using $\sum_{r=0}^{J-m}\kappa^r\le \sum_{r=0}^{J}\kappa^r$, we additionally obtain
$$
\|u^{\ell,\mathrm{mem}}(T_{\mathrm{off}}+m)\|_\infty
\leq\kappa^{m}\,\|u^{\ell,\mathrm{mem}}(T_{\mathrm{off}})\|_\infty,
$$
for all $0\le m\le J$.
\end{example}

Overall, our modelling choices (irrespective of the choice of decaying factors) are in line with the theory of memory mechanisms and, in particular, with~\cite{wickelgren1974strength}, where memory processes are divided into \emph{short-term} and \emph{long-term} traces.
{\color{black}{It is demonstrated that memory forms almost simultaneously with perception and decays in time. Short-term memory is proven to rapidly decay, and in our framework, such a rapid decay is formalized as an impulsive term. Conversely, long-term memory has been shown to exhibit a much slower decay of the memory trace. Therefore, we represent the persistent influence of long-term incentives as a normalized vanishing memory term.}}
At the same time, our choices imply that structural incentives modify agents' inherent biases that gradually decay but continue to shape decisions, while short-term interventions nudge adoption only temporarily.
\begin{remark}
     The balance between persistence and reactivity is governed by $\rho$ in \eqref{eq:input_eq}. Indeed, high values of $\rho$ emphasize structural inertia and long-term continuity, while low values increase responsiveness to instantaneous shocks.
\end{remark}
\begin{remark}
    The parameter $\rho$ and the sequence $\{\omega_j\}_j$ are characteristics of the social system and, thus, are not shaped by any external stimuli.
\end{remark}
\begin{remark} \label{rem:exp_omega}
    In this contribution, we will focus on $\omega_j$ defined as in Example~\ref{ex:IIR}, i.e., 
    we assume that the sequence of weights $\{\omega_j\}_{j \in \mathbb{N}_0}$ follows an exponentially decaying structure of the form
\begin{equation}\label{eq:exp_weights}
\omega_j = \left(1 - e^{-1/\tau}\right)e^{-j/\tau}, \qquad \tau > 0,
\end{equation}
where $\tau$ denotes the characteristic time scale governing the persistence of memory. This formulation ensures that the weights are positive, monotonically decreasing, and  $\sum_{j=0}^{\infty} \omega_j = 1$. This choice is also in line with the theory of memory mechanisms~\citep{wickelgren1974strength}, where it is shown that the long-term memory exhibits an exponential decay of the memory trace. In particular, it implies that  
the following recursive update with $\gamma \in [0, 1)$ holds:
\begin{equation} \label{eq:exp_decay} 
\omega_j = \gamma \cdot \omega_{j-1} \quad \text{for } j \geq 1, \mbox{ with } \gamma = e^{-1/\tau}.
\end{equation}
\end{remark}

\subsection{Augmented State Model with Memory Dynamics}
Let us now replace~\eqref{eq:exp_decay} in~\eqref{eq:mem_input}. This operation yields a first-order state equation for the memory component, i.e.,
\begin{equation}\label{eq:first_diff}
   u^{\ell, \text{mem}}(t) = \omega_0  u^\ell(t-1) + \sum_{j=1}^{t -1 } (\gamma\cdot \omega_{j-1})  u^\ell(t - j -1). 
\end{equation}
Through a change of variables (namely, setting $k=j-1$) and defining 
\begin{equation*}
    u^{\ell, \text{mem}}(t-1)=\sum_{k=0}^{t-1} \omega_{k} u^\ell(t - k -2), 
\end{equation*}
\eqref{eq:first_diff} is equivalent to
\begin{equation}
 u^{\ell, \text{mem}}(t)= \omega_0 \,  u^\ell(t-1) + \gamma \,  u^{\ell, \text{mem}}(t-1).
\end{equation}
These steps allow us to define the augmented state 
\begin{equation}
    \tilde{x}(t) = \begin{bmatrix} x(t) \\ u^{\ell, \text{mem}}(t) \end{bmatrix},
\end{equation}
which can be easily proven to evolve according to the augmented dynamics
\begin{subequations}
    \begin{equation}
        \tilde{x}(t\!+\!1)\! =\! \mathbf{A}_{\text{aug}} \tilde{x}(t) \!+\! \mathbf{B}_{\text{aug}}^{s} u^{s}(t)\!+\!\mathbf{B}_{\text{aug}}^{\ell} u^{\ell}(t)\!+\!\mathbf{B}_{\text{aug}}^{o} u^{o},
    \end{equation}
where
\begin{equation}\label{eq:A_aug}
\mathbf{A}_{\text{aug}}=\begin{bmatrix} \Lambda P & (I-\Lambda) \rho \\ 0 &\gamma {\color{black}{I}} \end{bmatrix},
\end{equation}
and 
\begin{equation}
\begin{aligned}
 \mathbf{B}_{\text{aug}}^{s}\!=\!\!&\begin{bmatrix} (I\!-\!\Lambda)(\mathbb{1}\!-\! \rho)\\ 0 \end{bmatrix},~\mathbf{B}_{\text{aug}}^{\ell}\!=\!\!\begin{bmatrix}
(I \!-\! \Lambda)\\
 \omega_0 I
\end{bmatrix} \\
& \qquad \qquad \qquad \mathbf{B}_{\text{aug}}^{o}\!=\!\!\begin{bmatrix}
(I - \Lambda)\\
0 
\end{bmatrix}\!. 
\end{aligned}
\end{equation}
\end{subequations}
Note that, since ${A}_{\text{aug}}$ is block triangular and $\gamma \in [0,1)$, then all its eigenvalues are strictly inside the unit circle when the spectral radius of $\Lambda P$ is smaller than one. In this case, the augmented system is Schur stable. In the remainder of the paper, we assume this condition on $\Lambda P$ to be satisfied.

\section{Model Properties} \label{sec:properties}
In the following, we present how the persistence of the incentive effects depends on the available budget and time horizon.
We begin with an \textit{ideal} and optimistic case in which the budget is assumed to be infinite. In this setting, we show that the effects of both the short-term and long-term incentive components are asymptotically preserved.
We then move to a more \emph{realistic} case in which the budget is finite and the time horizon is limited. In this case, the influence of the short-term incentive $u^s$ vanishes instantly, while the long-term effect persists. Finally, when considering a finite-budget setting with an infinite time horizon, the system converges asymptotically to the steady-state of the original control-free FJ model, demonstrating that the effect of both short- and long-term inputs vanishes asymptotically when a limited budget is considered, according to the realistic effect of policies (see the discussion in~\cite{callander2017dynamic}). \textcolor{black}{To this end, policies $\{u^\ell(t)\}_{t \in \mathbb{N}_0}$ are required to satisfy the following design constraint.}
\begin{design} \label{des:u_limits}
Nudging policies $u^\ell \text{ and } u^s$ are non-negative and must be designed satisfying
\begin{equation}
    \mathbb{0} \le \rho \odot u^{\ell,\text{mem}}(t) + (\mathbb{1} - \rho) \odot u^s(t) \leq \mathbb{1} - u^o \leq \mathbb{1} 
    \label{eq:policy_constraint}
\end{equation}
for all $t \in \mathbb{N}_0$ in order to ensure that  $ u(t)\in [0,1]^N$ and $x(t) \in [0,1]$ for all $t \in \mathbb{N}_0$.
\end{design}

\textcolor{black}{Note that this constraint will then be enforced in the control scheme proposed in Section~\ref{sec:control}.} 
Moreover, we assume that the policies to be designed are subject to resource limitations, as stated in the following.
\begin{assumption}\label{ass:budget}
Nudging policies operate under a fixed and finite budget $\beta \in \mathbb{R}_+$, with $\beta << \infty$, which is progressively depleted over time. Thus, the available resources for interventions at time $t \in \mathbb{N}_0$ are given by
\begin{subequations}
    \begin{equation}\label{eq:budget}
    U(t) = \beta - u^{\$}(t),
\end{equation}
where 
\begin{equation}\label{eq:u_dollaro}
    u^{\$}(t)=\sum_{k=0}^{t-1} \sum_{v \in V} \alpha u_{v}^{s}(t-k) + (1 - \alpha)u_{v}^{\ell}(t-k),
\end{equation}
\end{subequations}
and the parameter $\alpha \in [0,1]$ defines the relative conversion factor between the two control inputs in terms of their equivalent monetary value.
\end{assumption}
Last but not least, it is unreasonable for the budget available to nudge adoption to become negative. Hence, the following constraint is formalized.
\begin{design}\label{des:budget}
The available budget $U(t)$ is a non-negative quantity at each time step, i.e.,
$    U(t) \geq 0, \forall t \in \mathbb{N}_0,
$
ensuring that no intervention can be designed once the allocated resources are fully depleted. 
\end{design}


Let us first analyze the steady-state behavior of the system when the control policies satisfy the following assumption.
\begin{assumption}\label{ass:input_asy}
The long-term control input, $u^{\ell}(t)$ and the short-term control input, $u^s(t)$, converge to a constant value, i.e.,
\begin{subequations}
    \begin{equation}
    \lim_{t \to \infty} u^{s}(t) = \bar{u}^{s},
    \qquad
    \lim_{t \to \infty} u^{\ell}(t) = \bar{u}^{\ell},
    \end{equation}
    with $\bar{u}^{s}, \bar{u}^{\ell} \in [0, 1]^{N}$.
\end{subequations}
\end{assumption}
We can then formalize the following result.
\begin{proposition}
Let Assumptions \ref{ass:input_asy} and \eqref{eq:exp_weights} hold. Then, the asymptotic value of the latent inclinations under constant policies and infinite budget satisfies
\begin{subequations}
\begin{equation}
    x_{\infty} := \lim_{t\rightarrow\infty} x(t) =  (I-\Lambda P)^{-1}(I-\Lambda)u_{\infty},
\end{equation}
with $x_{\infty} \in [0, 1]^N$, and 
\begin{equation}
    u_{\infty} := \lim_{t\rightarrow\infty} u(t) =  u^o+ \rho \odot \bar{u}^\ell + (\mathbb{1} - \rho) \odot \bar{u}^{s},
\end{equation}
\end{subequations}
 with $u_{\infty}, \bar{u}^\ell, \bar{u}^s \in [0, 1]^N$.
\end{proposition}
\begin{pf}
The equilibrium state $x_{\infty}$ is trivially found by setting $x(t+1) = x(t) = x_{\infty}$ and $u(t) = u_{\infty}$ in~\eqref{eq:dyn_model}, i.e., 
\begin{subequations}
\begin{equation} \label{eq:equil_asy}
x_{\infty} = (I - \Lambda P)^{-1}(I-\Lambda)u_{\infty},
\end{equation}
with $u_{\infty}:= \lim_{t\rightarrow\infty} u(t)$ being
\begin{align*}
    u_{\infty} & = u^{o} \!+ \rho\! \odot\! \lim_{t\rightarrow\infty} u^{\ell,\text{mem}}(t) + (\mathbb{1} \!-\! \rho) \!\odot\! \lim_{t\rightarrow\infty} u^s(t)\\
    &=u^{o} \!+ \rho\! \odot\! \lim_{t\rightarrow\infty} u^{\ell,\text{mem}}(t) + (\mathbb{1} \!-\! \rho) \!\odot\!\bar{u}^{s},
\end{align*}
where the last equality holds based on Assumption~\ref{ass:input_asy}. Meanwhile, according to~\eqref{eq:mem_input} and Assumption~\ref{ass:input_asy}, $u^{\ell,\text{mem}}(t)$ satisfies 
\begin{equation}
     \lim_{t\rightarrow\infty} u^{\ell,\text{mem}}\!=\! \lim_{t\rightarrow\infty}\!\sum_{j=0}^{t-1} \omega_j u^\ell(t - j -1)=\!\sum_{j=0}^{\infty} \omega_j  \bar{u}^\ell.
\end{equation}
As the sum of the weights satisfies \eqref{eq:weight_prop}, it further holds that
\begin{equation}
    \lim_{t\rightarrow\infty} u^{\ell,\text{mem}}(t) = \bar{u}^\ell  \sum_{j=0}^{\infty}\omega_{j} = \bar{u}^\ell \cdot 1 = \bar{u}^\ell.
\end{equation}
Therefore, since $\bar{u}^\ell \in [0,1]^N$, $\lim_{t\rightarrow\infty} u^{\ell,\text{mem}}(t) \in [0,1]^N$, and  
    \begin{equation}
    x_{\infty} = (I - \Lambda P)^{-1}(I-\Lambda)u_{\infty},
    \end{equation}
    with
    \begin{equation}
    u_{\infty} = u^{o} + \rho \odot \bar{u}^\ell + (\mathbb{1} - \rho) \odot \bar{u}^{s},
    \end{equation}
\end{subequations}
thus concluding the proof. \hfill $\square$
\end{pf}

We now analyze how past long-term and short-term control inputs persist in the system state, considering a scenario where constant policies are applied for a finite number of time instants $T_B$. For a realistic analysis of the effect of incentives, we will analyze the behavior of the system for a finite time horizon $t \in \mathbb{N}_0$, with $t \ll \infty$.
\begin{assumption}
\label{ass:limited_incentives}
Let $T_B \ll \infty$ be the last time instant at which a non-zero control inputs are applied (e.g., due to budget depleting), i.e.,
\begin{equation*} \label{eq:cessation}
    u^\ell(t) \neq 0, \quad u^s(t) \neq 0 \quad \text{for } t \leq T_B,
\end{equation*}
\begin{equation*}
    u^\ell(t) = 0, \quad u^s(t) = 0 \quad \text{for } t > T_B.
\end{equation*}
\end{assumption}
Under this assumption, we can formalize the following property.
\begin{proposition}
Let Assumption~\ref{ass:limited_incentives} hold. Then, the value of the latent inclinations for $t > T_B, ~t \ll \infty$ is 
\begin{align}\label{eq:results_prop2}
    \nonumber x(t+1) &= \Lambda P x(t) + (I-\Lambda) u(t) \\
           &= \Lambda P x(t) + (I-\Lambda) \left( u^o + \rho \odot u^{\ell,\text{mem}}(t) \right).
\end{align}    
\end{proposition}
\begin{pf}
While $u^s(t)$ on the current state evolution is immediately zero for $t > T_B$, the long-term input $u^\ell(t)$ affects the state dynamics through the normalized, weighted-average term $u^{\ell,\text{mem}}(t)$ defined in~\eqref{eq:mem_input}. The summation in~\eqref{eq:mem_input} can be split into two terms:
\begin{align*}
    u^{\ell,\text{mem}}(t) &=\!\! \sum_{j=0}^{t-T_B} \!\omega_j(t) u^\ell(t\! -\! j\! -\!1)\!+\!\!\!\!\!\sum_{j=t-T_B+1}^{t-1}\!\!\!\!\omega_j(t)u^\ell(t\! -\! j\! -\!1), \\
\end{align*}
where the first term comprises all null inputs, while the second one contains all non-zero ones.
Accordingly, based on \eqref{eq:dyn_model} and \eqref{eq:input_eq}, it is straightforward to prove that the state dynamics for $t > T_B$ becomes the one in \eqref{eq:results_prop2}, thus concluding the proof.  \hfill $\square$
\end{pf}



We can lastly formalize the (asymptotic) vanishing effect of the long-term policy under limited budget as follows.  
\begin{proposition}
Let Assumption \ref{ass:limited_incentives} hold. Then, the asymptotic value of the latent inclinations under policies with a finite budget satisfies
\begin{equation}
    x_{\infty} := \lim_{t\rightarrow\infty} x(t) = (I-\Lambda P)^{-1}(I-\Lambda)u^o,
\end{equation}
that coincides with the open-loop asymptotic state and the FJ equilibrium.
\end{proposition}

\begin{pf}
It is straightforward that the equilibrium state $x_{\infty}$ of \eqref{eq:dyn_model} is
\begin{equation} \label{eq:equil_asy_lim_new}
x_{\infty} = (I - \Lambda P)^{-1}(I-\Lambda)u_{\infty},
\end{equation}
where $u_{\infty} := \lim_{t\rightarrow\infty} u(t)$ satisfies
\begin{align*}
    u_{\infty}&= u^{o} + \rho \odot \lim_{t\rightarrow\infty} u^{\ell,\text{mem}}(t) + (\mathbb{1} - \rho) \odot \lim_{t\rightarrow\infty} u^s(t)\\
    &=u^{o}+ \rho \odot \lim_{t\rightarrow\infty} u^{\ell,\text{mem}}(t),
\end{align*}
where the last equality holds thanks to Assumption \ref{ass:limited_incentives}. Meanwhile, letting $T_B$ be the instant when the budget is depleted, according to the same Assumption, then $u^{\ell,\text{mem}}(t)$ can be rewritten as
\begin{align*}
    u^{\ell,\text{mem}}(t) &=
    \sum_{j=t-T_B+1}^{t-1}\!\omega_j {u}^{\ell}(t\!-\!j\! -\!1). 
\end{align*}
Since $\omega_j$ is exponentially vanishing, the contribution of the nonzero terms vanishes, and thus $u^{\ell,\text{mem}}(t)$ vanishes. Substituting these limits into $u_{\infty}$ concludes the proof. \hfill $\square$
\end{pf}

\section{Control Strategies to foster sustainable behaviors} \label{sec:control}
 Using the model presented in Section~\ref{sec:model}, we now propose two possible strategies that policymakers can employ to design policies that nudge the adoption of a sustainable behavior over a finite time frame of length $T$. In designing them, we assume that the initial inherent predisposition $u^o$ (or its estimate, e.g., inferred from survey data as in~\cite{Villa2024}) is available for policy design. 

 We first introduce a na\"ive strategy for policy design, in which the available budget $\beta$ is equally distributed across agents and over time, giving each of them the same constant long-term and short-term inputs $\bar{u}^s$ and $\bar{u}^\ell$. Note that for these constant values to lie in $[0,1]^{N}$ (and, thus, the resource allocation to be feasible), we saturate $\beta = \min(\beta,\, TN)$.
 \begin{policy}[A simple distributive policy]
     The constant policy value is defined as 
     \begin{equation}
         \bar{u} = \frac{\beta}{TN},~~~\text{with } 0 \leq \bar{u} \leq 1.
     \end{equation}
    Since the maximum short-term effort to be applied without saturating the system dynamics is
    \begin{equation}
        u^s_{i,\mathrm{max}}  = \min\!\left(1,\, \max\!\left(0,\,\frac{1 - u^{o}_i - \rho_i}{1 - \rho_i}\right)\right),~~i\!=\!1,\ldots,N,
    \end{equation}
    we then define the $i$-th component of $\bar{u}^s$ as 
    \begin{equation}
        \bar{u}^s_{i}  = \min\!\left({\bar{u}},\, u^s_{i,\mathrm{max}} \right),
    \end{equation}
    for $i=1,\ldots,N$. The long-term policy is instead given by $\bar{u}^\ell_{i} = \bar{u},~ i=1,\ldots,N.$
 \end{policy}

 Note that this policy can be completely designed beforehand. Instead, let us assume that the policymaker can monitor the evolution of opinions and dynamically adjust their strategies, which is possible if the policy design steps are relatively infrequent (e.g., every 6 months), as is common in policy design for innovation diffusion. 
 \begin{policy}[MPC for opinion nudging]
 In this case, the resource allocation problem can be formulated as a constrained, model-predictive control problem over a prediction horizon $L\geq 1$ to be solved in a receding horizon fashion at each time $t \in \mathbb{N}_0$, i.e., 
\begin{subequations}\label{eq:MPC_Mixed_Memory}
    \begin{align}
        & \underset{\mathcal{X}_{|t},~\mathcal{U}_{|t}^{s},~\mathcal{U}_{|t}^{\ell}}{\min}~J^{\mathrm{MPC}}(\mathcal{X}_{|t},\mathcal{U}_{|t}^{s},\mathcal{U}_{|t}^{\ell}),\\
        & \mbox{s.t.} \quad~ x_{|t}(h\!+\!1)=\Lambda Px_{|t}(h)+(I-\Lambda)u_{|t}(h),\\
        & \qquad~~ u_{|t}(h)\!=\!u^o \!+\! \rho \odot u^{\ell, \text{mem}}_{|t}(h) \!+\! (\mathbb{1}\!-\!\rho) \odot u_{|t}^s(h),\\
        & \quad~~\quad u^{\ell, \text{mem}}_{|t}(h+1) = \gamma\, u^{\ell, \text{mem}}_{|t}(h) + \omega_0\, u_{|t}^\ell(h)),\\
        & \qquad~~
u_{|t}^{\$}(h)=\sum_{\kappa=0}^{h}~\sum_{v \in \mathcal{V}} \alpha u_{|t ,v}^{s}(\kappa)+(1-\alpha)u_{|t ,v}^{\ell}(\kappa),\label{eq:budg_def_MPC}\\
        & \qquad~~u_{|t}^{\$}(h)\leq U(t)\!-\!u_{|t}^{\$}(h\!-\!1),\label{eq:budget_constr_MPC}\\ 
        &\qquad~~u_{|t}^{s}(h), u_{|t}^{\ell}(h), u_{|t}(h) \in [0,1]^N,~h \!\in\! [0,L-1],\\
        &\qquad~~x_{|t}(0)\!=\!x(t),u^{\ell, \text{mem}}_{|t}(0)\!=\!u^{\ell, \text{mem}}(t),
    \end{align}
where $\mathcal{X}_{|t}\!=\!\{x_{|t}(h)\}_{h=0}^{L}$, $\mathcal{U}_{|t}^{s}\!=\!\{u_{|t}^{s}(h)\}_{h=0}^{L-1}$, $\mathcal{U}_{|t}^{\ell}\!=\!\{u_{|t}^{\ell}(h)\}_{h=0}^{L-1}$, $U(t)$ and $u_{|t}^{\$}$ are defined as in Assumption~\ref{ass:budget}. The loss is defined as 
\begin{align}
    \nonumber &J^{\mathrm{MPC}}(\mathcal{M}_{|t},\,\mathcal{U}_{|t}^{\ell},\,\mathcal{U}_{|t}^{s})=\sum_{h=0}^{L-1} \left[\|\mathbb{1}\!-\!x_{|t}(h)\|_{Q}^{2}+\|u^{s}_{|t}(h)\|_{R_1}^{2}\right.\\
    &\qquad\qquad \qquad  \left. +\|u^{\ell}_{|t}(h)\|_{R_2}^{2}\right]\!+\!V_f\big(x_{|t}(L),\, u^{\ell,\text{mem}}_{|t}(L)\big),
\end{align}
where $Q \succ 0$ and $R_i\succ 0$, with $i=1,2$, control the trade-off between the widespread adoption of sustainable behavior and the long-term and short-term policy efforts needed to achieve it. 
\end{subequations}
Meanwhile, the terminal cost $V_f\big(x_{|t}(L),\, u^{\ell,\text{mem}}_{|t}(L)\big)$ is selected as
\begin{align}
    \nonumber & V_f\big(x_{|t}(L),\, u^{\ell,\text{mem}}_{|t}(L)\big)\!=\!\|\mathbb{1} - x_{|t}(L)\|_{P_{11}}^2\!\!+\!\|u^{\ell,\text{mem}}_{|t}(L)\|_{P_{22}}^2
    \\ & ~~~~~~~~~~~~~~~~~~~~~~~~~~+ 2\,(\mathbb{1} -  x_{|t}(L)^{\top}) P_{12} u^{\ell,\text{mem}}_{|t}(L),    
\end{align}
where $P_{11}$, $P_{22}$ and $P_{12}$ are the diagonal and counter-diagonal blocks of the symmetric matrix $P_L \succ 0$ satisfying 
\begin{equation}
    A_{\text{aug}}^{\top}  P_L A_{\text{aug}}- P_L = - Q_{L},
\end{equation}
and $A_{\text{aug}}$ defined as in\footnote{In our example, $\gamma=e^{-1/\tau}$.} \eqref{eq:A_aug}. Note that, according to our assumptions $A_{\text{aug}}$ is Schur stable and, thus, this choice guarantees the stability of the closed-loop system (see~\cite[Chapter 2]{rawlings2020model}). 
 \end{policy}
\begin{remark}\label{remark:manifest_op}
    While this MPC scheme relies on the knowledge of $x(t)$, policymakers are likely not to have access to the actual opinion of the individuals, but rather to the binary $ \{y(t)\}_{t \in \mathbb{N}_0}$ which can be modelled, e.g., as a Bernoully with mean $x(t)$ (see~\cite{ravazzi2025optimal}). This binary evidence can be used to initialize $x_{|t}(0)$ in \eqref{eq:MPC_Mixed_Memory}, e.g., by recursively computing the temporal mean of $\{y(k)\}_{k=0}^{t}$, implementing a leaky integrator, or applying any other sliding-window estimator. 
\end{remark}


\section{A simulation example on real data} \label{sec:num_example}
\begin{figure}[!tb]
    \centering
    \includegraphics[width=0.8\linewidth]{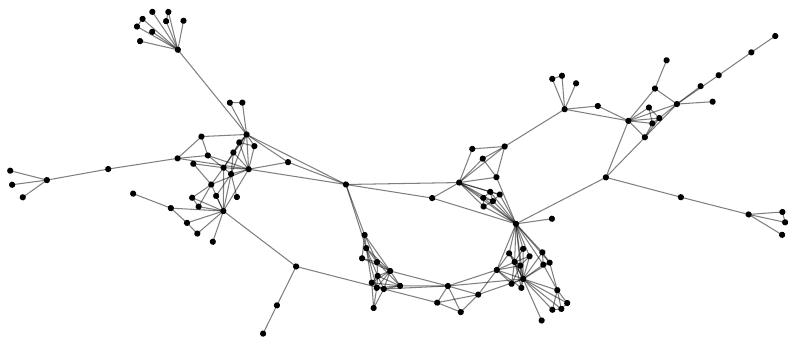}
    \caption{Undirected graph of $N=112$ agents.}
    \label{fig:und_network}
\end{figure}
We now analyze how policies designed with the strategy introduced in Section~\ref{sec:control} impact opinion formation and adoption dynamics. To this end, we consider the survey on sustainable mobility habits in~\cite{fiorello2015eu}, previously used in~\cite{Villa2024} to assess alternative policy design strategies for nudging virtuous behaviors. As in the latter work, we focus on a subset of $N=112$ respondents living in the metropolitan area of Milan and connected according to a proxy of their geographical proximity inferred from individual mobility habits as depicted in \figurename{~\ref{fig:und_network}}. Note that, as in~\cite{Villa2024}, we assume that the connections among agents are shaped by the agent’s credibility $c_v$ for all $v \in \mathcal{V}$, which reflects how socially trusted each agent is in the network. The latter is calculated for each agent starting from their reliability $\zeta_v$, which assumes a value between $0$ and $1$ depending on the educational level of the respondent (used as a proxy of perceived competence), having it every time the $v$-th individual belongs to one or more social groups that are subject to prejudice (e.g., youth, women, or low-income individuals). After computing individuals' credibility, we construct the  credibility matrix $ \mathbf{C} =  \begin{bmatrix} c_1 \mathbb{1} & c_2 \mathbb{1} & \dots & c_N \mathbb{1}
\end{bmatrix} ~ \in \mathbb{R}^{N \times N}$, and each row is normalized, obtaining the row-stochastic matrix $\text{C}$. The resulting influence matrix is then defined as $P = \mathcal{A} \cdot \text{C}$, with $\mathcal{A}$ denoting the adjacency matrix of the graph depicted in \figurename{~\ref{fig:und_network}}.    

In addition, we use the individual’s inherent reluctance to adopt electric vehicles (EVs) computed in~\cite{Villa2024} to estimate the inherent bias of each agent. In particular, with a slight abuse of notation to be aligned with \cite{Villa2024}, we set $u^o_v = \mathbb{1} - \rho_v(0)$ for all $v \in \mathcal{V}$, where $\rho_v(0)$ is estimated reluctance to adoption for the $v$-th agent. Finally, in line with what is done in~\cite{Villa2024}, we randomly assign the receptivity to policies to each agent $(1 - \lambda_v)$, as no information is provided on the survey to estimate even a proxy for it. 

\begin{table}[!tb]
    \centering
    \caption{Parameters for simulation and MPC design.}
    \label{tab:parameters}
    \begin{tabular}{cccccc}
         $\tau$ \eqref{eq:exp_decay} & $L$ & $R_{1}$ & $R_{2}$ & $Q$ & $Q_L$ \\
         \hline
         3 & 10 & $10I$ & $10I$ & $100I$ & $I$\\
         \hline 
    \end{tabular}
\end{table}
With these choices, our simulations are performed over $T=11$ steps by considering the model parameters as well as the design choices for the MPC strategy reported in \tablename{~\ref{tab:parameters}}. In all simulation, we assume the actual individual inclinations not to be accessible and, in line with Remark~\ref{remark:manifest_op}, the mean inclination is estimated from the binary evidence of adoption as $\mu(t) = \frac{1}{t+1} \sum_{\tau = 0}^{t} y(\tau)$. This estimate, which does not require us to tune any additional hyperparameter, is then used to initialize the state in the MPC scheme presented in Section~\ref{sec:control}. Note that, even if this estimate can be biased, the considered estimation window is short enough for such a bias to be negligible.   

\subsection{Simulation results}
\begin{figure}[!tb]
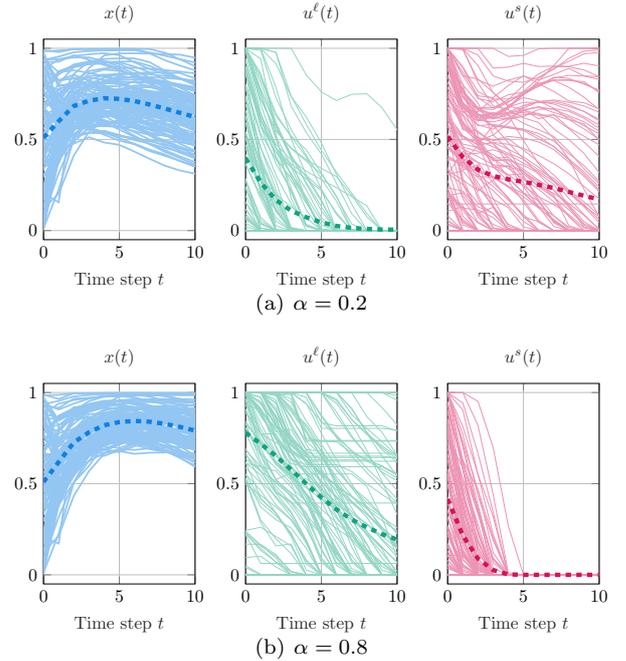

\centering
\begin{tabular}{c}
    \subfigure[$\alpha = 0.2$]{%
        \label{fig:alpha02rho07bud200}
        \resizebox{0.9\linewidth}{!}{\input{Figures/alpha02rho07bud200}}
    } \\[0.2cm]
    \subfigure[$\alpha = 0.8$]{%
        \label{fig:alpha08rho07bud200}
        \resizebox{0.9\linewidth}{!}{\input{Figures/alpha08rho07bud200}}
    }
\end{tabular}\vspace{-.2cm}
\caption{$x(t)$, $u^\ell(t)$ and $u^s(t)$, varying $\alpha$ for $\rho = 0.7$ and $\beta = 200$. The dashed lines represent the mean across agents.}
\label{fig:alpha_rho07_bud200}
\end{figure}

Let us first consider two possible cases for $\alpha$ in \eqref{eq:u_dollaro}, namely $\alpha = 0.2$ and $\alpha = 0.8$, while fixing $\rho = 0.7$ and $\beta = 200$. As shown in \figurename{~\ref{fig:alpha02rho07bud200}} and \figurename{~\ref{fig:alpha08rho07bud200}}, when $\alpha$ is small, the controller allocates more effort to short-term interventions, leading to faster but less sustained effects on the acceptance.  
Conversely, for larger $\alpha$, the system privileges the long-term component, accumulating a stronger memory effect. Indeed, as shown in \figurename{~\ref{fig:alpha08rho07bud200}}, the acceptance variable reaches higher steady-state values, closer to $1$, compared to \figurename{~\ref{fig:alpha02rho07bud200}}.

\begin{figure}[!tb]
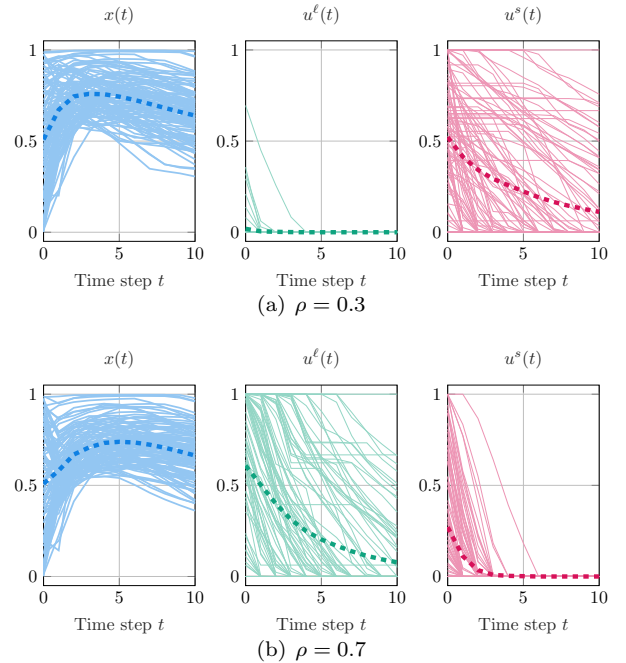

\centering
\begin{tabular}{c}
    \subfigure[$\rho = 0.3$]{%
        \label{fig:alpha05rho03bud200}
        \resizebox{0.9\linewidth}{!}{\input{Figures/alpha05rho03bud200}}
    } \\[0.2cm]
    \subfigure[$\rho = 0.7$]{%
        \label{fig:alpha05rho07bud200}
        \resizebox{0.9\linewidth}{!}{\input{Figures/alpha05rho07bud200}}
    }
\end{tabular}\vspace{-.2cm}
\caption{$x(t)$, $u^\ell(t)$, and $u^s(t)$, varying $\rho$, for $\alpha = 0.5$ and $\beta = 200$. The dashed lines represent the mean across agents.}
\label{fig:alpha_rho07_bud200}
\end{figure}

\figurename{~\ref{fig:alpha05rho03bud200}} and \figurename{~\ref{fig:alpha05rho07bud200}} instead illustrate the effect of $\rho$ when fixing $\alpha=0.5$. Specifically, we consider $\rho=0.3$ and $\rho=0.7$. Our results show that, when $\rho=0.3$, the controller relies more on short-term actions with respect to the case in which $\rho=0.7$, especially at the beginning of the horizon, resulting in a slower increase in acceptance.  Moreover, with $\rho = 0.3$, the residual budget at the final time step is approximately $42$, whereas for $\rho = 0.7$ it is around $15$.  This confirms that a stronger short-term weight leads the system to adopt a more parsimonious spending strategy, deferring resource consumption within the MPC prediction horizon.

\begin{figure}[!tb]
\centering
    \resizebox{0.9\linewidth}{!}{\input{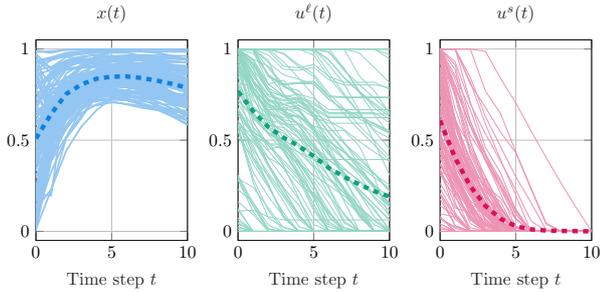}}\vspace{-.2cm}
    \caption{Increased budget: $x(t), ~u^\ell(t), ~u^s(t)$ for $\alpha=0.5$, $\rho=0.7$, and $\beta=400$. The dashed lines represent the mean over the agents.}
    \label{fig:alpha05rho07bud400}
\end{figure}

Finally, comparing \figurename{~\ref{fig:alpha05rho07bud200}} and \figurename{~\ref{fig:alpha05rho07bud400}}, for $\alpha = 0.5$ and $\rho = 0.7$, while varying the budget (we consider $\beta = 200$ and $\beta = 400$ respectively), it is evident that increasing the available resources leads to higher acceptance.

\subsection{Na\"ive vs Receding Horizon policy}

\begin{table}[!tb]
\centering
\caption{Na\"ive vs Receding Horizon (RH) policy. ${\sigma_x(T)}$ is the standard deviation of opinions at the last time instant, while $r_\beta$ is the unused budget.}
\label{tab:budget_results}
\begin{tabular}{lcccccc}
{Policy} & ${\bar{x}(T)}$ & ${\sigma_x(T)}$ & ${{u}^{s}_{\text{mean}}}$ & ${{u}^{\ell}_{\text{mean}}}$ & $\beta$ &  $r_\beta$\\
\midrule
RH & 0.77 & 0.11 & 0.13 & 0.42 & 400 & 57.33 \\
Const. & 0.67 & 0.11 & 0.08 & 0.32 & 400 & 148.31 \\
\midrule
RH & 0.67 & 0.14 & 0.04 & 0.26 & 200 & 16.74 \\
Const. & 0.55 & 0.12 & 0.05 & 0.16 & 200 & 70.29 \\
\bottomrule
\end{tabular}
\end{table}

We then compare the two policies discussed in Section~\ref{sec:control}, fixing $\alpha=0.5$, and $\rho=0.7$. As shown in \tablename{~\ref{tab:budget_results}}, the RH policy consistently achieves a higher final average state $\bar{x}(T)$ than the na\"ive controller for both budget levels considered (i.e., $\beta=200$ and $\beta=400$). Moreover, for both budget levels, RH uses almost all available resources, whereas constant control leaves a significant fraction unused.
For both cases, the mean values of $u^s$ and $u^\ell$ over time and across agents (${u}^{s}_{\text{mean}}$ and ${u}^{\ell}_{\text{mean}}$) indicate that both policies design methods allocate more effort to long-term control. In general, it can be observed that the RH policy achieves better performance by prioritizing interventions where they are most effective.

\section{Conclusions} \label{sec:conclusions}
In this work, we propose an opinion dynamics model that captures both short- and long-term behavioral shifts induced by external policies. We do so by introducing memory in the long-term effect of incentives, which impacts individual opinion dynamics. We then leverage this model to propose two policy design strategies under budget constraints. We evaluate the impact of such policies on opinion dynamics through numerical simulations based on real data. Future work will focus on studying the persistence of opinions under varying memory lengths and network structures, exploring finite impulse response (FIR) weighting, and focusing on estimation methods for individuals’ predispositions.

\bibliography{final_submitted}             
\end{document}